\documentclass[preprint,12pt,authoryear,review]{elsarticle}

\usepackage{amssymb}
\usepackage{amsmath}
\usepackage{graphicx}
\usepackage{multirow}
\usepackage{array}
\usepackage{float}
\usepackage{rotfloat}
\usepackage{nameref}
\usepackage{makecell}
\newcommand{\mm}[1]{\mathrm{#1}} 

\journal{Earth and Planetary Science Letters}

\begin{document}

\begin{frontmatter}



\title{A Collective Trigger for Widespread Planetesimal Formation Revealed by Accretion Ages} 


\author[label1]{James F J Bryson} 
\author[label1,label5]{Hannah R Sanderson} 
\author[label2]{Francis Nimmo} 
\author[label1,label6]{Sanjana Sridhar} 
\author[label4]{Gregory A Brennecka} 
\author[label3]{Yves Marrocchi} 
\author[label1]{Jason P Terry} 

\affiliation[label1]{organization={Department of Earth Sciences, University of Oxford},
            addressline={South Parks Road}, 
            city={Oxford},
            postcode={OX2 8FW}, 
            country={UK}}
\affiliation[label5]{organization={Centre for Planetary Habitability, Department of Geosciences, University of Oslo},
            city={Oslo}, 
            country={Norway}}
\affiliation[label2]{organization={Department of Earth and Planetary Sciences, University of California Santa Cruz},
            city={Santa Cruz},
            postcode={95060}, 
            state={California},
            country={USA}}
\affiliation[label6]{organization={Laboratoire Magmas et Volcans, Université Clermont Auvergne. CNRS. IRD. OPGC},
            city={Clermont-Ferrand},
            country={France}}
\affiliation[label4]{organization={Nuclear and Chemical Science Division, Lawrence Livermore National Laboratory},
            city={Livermore},
            postcode={94550}, 
            state={California},
            country={USA}}
\affiliation[label3]{organization={Centre de Recherches Pétrographiques et Géochimiques (CRPG), CNRS, UMR 7358},
            city={Nancy},
            country={France}}

\begin{abstract}
The formation of planetesimals was an integral part of the cascading series of processes that built the terrestrial planets. To illuminate planetesimal formation, here we develop a refined thermal evolution model to calculate the formation ages of meteorite parent planetesimals. This model includes chemical reactions and phase changes during heating, as well as natural variations in the proportions of the constituent phases of these planetesimals. We find that the parent bodies of non-carbonaceous (NC) and carbonaceous (CC) iron meteorites start forming at very similar times ($\sim$0.95 Myr after calcium-aluminium-rich inclusion [CAI] formation) and occupy overlapping time windows. NC and CC chondrite parent bodies formed later during non-overlapping periods. We combine these ages with proportions of isotopic end-members we recover from mixing models to construct records of motion throughout the protoplanetary disk. These records argue that NC and CC material traversed the barrier in the disk after $\sim$0.95 Myr after CAI formation. The onset of this motion coincided with planetesimal formation, indicating that the phenomenon that drove motion also triggered planetesimal formation. We argue that this feature also served as the semi-permeable barrier in the disk. Although its identity is uncertain, the effects this phenomenon had on the timing of planetesimal formation and motion through the disk can now serve as constraints on models of disk evolution. Models that reproduce these effects would elucidate the nature and implications of this phenomenon, which is key to unlocking a holistic model of terrestrial planet building.

\end{abstract}

\begin{highlights}
\item We calculate formation ages of the parent planetesimals of a suite of meteorites
\item NC and CC planetesimals start forming at very similar times and occupy overlapping periods
\item These ages indicate that both NC and CC material traversed the barrier in our disk
\item NC and CC planetesimal formation was triggered by a defining feature/event
\item This feature/event also drove isotopic evolution throughout the protoplanetary disk
\end{highlights}


\end{frontmatter}


\section{Introduction}
\label{Introduction}

The terrestrial planets most likely formed through the collision and merging of millions of planetesimals \citep{Morbidelli:2025p119120}. The principal parameters that governed the properties of these ancient, $\sim$10-1000 km-sized planetary bodies were their formation times and formation processes \citep{ Dodds:2021pe2020JE006704,Sanderson:2024p119083}. Thus, uncovering a complete understanding of these factors will help to elucidate the nature and evolution of our own planet. 

Formation time is key to the behaviour of a planetesimal because it sets the amount of radioactive $^{26}$Al it accreted \citep{Merk:2002p183,Hevey:2006p95,Neumann:2012pA141}. This time therefore controlled the planetesimal's peak metamorphic temperature (PMT), and so determined whether the planetesimal melted and differentiated, and when these processes would have occurred. Fragments of planetesimals arrive on Earth as meteorites, and measurements of these extraterrestrial rocks have revealed the PMTs (e.g., \citealt{Huss:2006,VerdierPaoletti:2017p273}) and differentiation times (e.g., \citealt{Hellmann:2024p118518,Spitzer:2025P134}) of numerous planetesimals. These measurements can be translated into the formation ages of these meteorites' parent bodies using models of planetesimal thermal evolution. However, previous examples of such models have employed simplified approaches that do not consider the suite of chemical reactions and phase transitions that occur during planetesimal heating when calculating accretion ages \citep{Sugiura:2014p772,Bryson:2021p163,Spitzer:2021p117211}, meaning these times are comparatively poorly constrained. Moreover, significant progress has been made recently in refining the differentiation ages of planetesimals \citep{Spitzer:2021p117211,Hellmann:2024p118518}, meaning that previous formation ages require revision.

The process by which planetesimals formed is essential to their evolution because it determined their sizes and possible accretion locations within the protoplanetary disk. The dominant process believed to have formed planetesimals is the streaming instability \citep{Youdin:2005p459}, which requires a local increase in the dust-to-gas ratio in the disk. This increase can result from changes in the disk's pressure gradient (e.g., within a pressure maximum), which can be caused by features such as planets \citep{Lin:1993p749} and/or condensation lines \citep{Bryden:2000p1091}. The specific phenomena that promoted planetesimal formation in our solar system are currently unknown, limiting our understanding of this key stage of planet building. However, each candidate feature will have had a unique impact on the motion of gas, dust, and solids through the solar system \citep{Lin:1993p749,Drazkowska:2017pA92,Ziampras:2020pA29}. As such, records of dynamics in our disk would reveal key insights into the mechanism behind planetesimal formation. 

Material that originated from different regions of the disk exhibits distinct ratios of nucleosynthetic isotopic anomalies \citep{Bermingham:2020p133}, and planetesimals are built from varying mixtures of these materials \citep{Bryson:2021p163}. The bulk nucleosynthetic isotopic compositions of meteorites therefore reflect the motion of material throughout the disk, and so encode the records of dynamics that we require. Despite this potential, incomplete archives of nucleosynthetic isotopic compositions measured from meteorites coupled with inaccurate planetesimal accretion ages have hindered previous efforts to recover these records \citep{Kleine:2020p55,Bryson:2021p163}. However, recent advances in the availability and breadth of these isotopic compositions (e.g., \citealt{Anand:2021p6,Spitzer:2024peadp2426,Spitzer:2025P134}) mean more complete and reliable records of dynamics in the disk are now attainable.

To better constrain the formation times and processes of planetesimals, here we build and employ a refined planetesimal formation age model. We use this to calculate planetesimal accretion times from differentiation ages measured from iron meteorites and PMTs gleaned from chondrites. Alongside this, we use recent breakthroughs in measurements of nucleosynthetic isotopic compositions of a variety of meteorites to recover the most complete records yet of mixing in our disk. Combined, these chronicles of mixing and accretion yield records of motion throughout our disk. These records argue that a collective trigger most likely kickstarted planetesimal formation throughout the solar system. Although the identity of this trigger is currently unknown, it played a a crucial role in terrestrial planet building. As such, identifying and understanding this feature is key to unlocking a holistic model of the transformation of dust into planets.

\section{Methods}
\label{Methods}

\subsection{Isotopic Mixing Proportions}
\label{mix}
The proportions of the various components that make up meteorites reflect the mixing and motion of dust and solids in the protoplanetary disk. Because these components can have a distinct nucleosynthetic isotopic composition (e.g., \citealt{Bermingham:2020p133}), it is possible to calculate the proportion of each component in a meteorite from its bulk isotopic composition. Carbonaceous (CC) meteorite groups can be expressed as mixtures of material with CI (Ivuna-like), non-carbonaceous (NC), and calcium-aluminium-rich inclusion (CAI) isotopic compositions \citep{Bryson:2021p163,Yap:2023p115680}. Here, we take CI material to be dust with isotopic and chemical compositions that match CI chondrites. Because ureilites often represent an end-member within the NC reservoir \citep{Spitzer:2020pL2}, we take NC material to be dust with isotopic compositions that match ureilites (see Supplementary Information). We take CAI material to be refractory objects with isotopic and chemical compositions that match CAIs (i.e., CAIs themselves). The isotopic composition of a CC meteorite, \(I_{met}\), can then be written as
\begin{equation}
I_{\mm{met}}=\frac{I_{\mm{NC}}C_{\mm{NC}}P_{\mm{NC}}+I_{\mm{CI}}C_{\mm{CI}}P_{\mm{CI}}+I_{\mm{CAI}}C_{\mm{CAI}}P_{\mm{CAI}}}{C_{\mm{NC}}P_{\mm{NC}}+C_{\mm{CI}}P_{\mm{CI}}+C_{\mm{CAI}}P_{\mm{CAI}}},
\end{equation}
where \(I\), \(C\), and \(P\) denote the isotopic composition, chemical concentration of the isotope system $I$, and mass proportion of each end-member component, respectively. Using measured compositions of two elements whose isotopic signatures trace different components (e.g., $\varepsilon^{54}$Cr and $\varepsilon^{50}$Ti; \citealt{Bryson:2021p163}) and the condition that \(P_{\mm{NC}}+P_{\mm{CI}}+ P_{\mm{CAI}}=1\), we were able to formulate and solve a pair of simultaneous equations to ultimately yield \(P_{\mm{NC}}\), \(P_{\mm{CI}}\), and \(P_{\mm{CAI}}\) for each meteorite. Additionally, these isotopic proportions can be converted into the fractions of chondrules, matrix, and CAIs (see Supplementary Information) in chondrites as well as the chondritic progenitors to iron meteorites \citep{Bryson:2021p163}. 

The ability to express NC meteorites as mixtures of distinct materials from the protoplanetary disk is less well explored than it is for CC meteorites \citep{Alexander:2019p246}. However, in a plot of $\varepsilon^{54}$Cr and $\varepsilon^{50}$Ti, NC meteorites form a near-linear trend that extends to CC meteorites (Fig. S3). This behaviour argues that NC meteorites can be expressed as mixtures of an NC end-member (taken here as ureilites) and CC material. \cite{Williams:2020p05235} and \cite{Burkhardt:2021peabj7601} note that this trend may not extend through CI material, so NC meteorites may consist of different end-members to CC meteorites. However, CC material is itself an evolving mixture of NC, CI, and CAI material (see Section \ref{CCDiss}). As such, the CC material being added to the NC reservoir had a component of CI material within it. To be consistent with our description of CC meteorites, we adopted CI material as the CC end-member in our a mixing model of NC meteorites. 

Due to the metal-rich compositions of iron meteorites and the refractory nature of CAIs, we adopted different approaches to calculate \(P_{\mm{NC}}\), \(P_{\mm{CI}}\), and \(P_{\mm{CAI}}\) from different meteorite types:
\begin{itemize}
    \item \textit{CC iron meteorites---IIC, IID, IIF IIIF, IVB, ungrouped.} We calculated the proportions of isotopic end-members for these meteorites using their bulk $\varepsilon^{62}$Ni and $\Delta^{95}$Mo values \citep{Dauphas:2024p115805,Spitzer:2025P134,Spitzer:2025p119530}. Among iron meteorites, $\Delta^{95}$Mo primarily reflects \(P_{\mm{CAI}}\) while $\varepsilon^{62}$Ni predominately reflects \(P_{\mm{NC}}\) and \(P_{\mm{CI}}\) \citep{Bryson:2021p163}. 
    \item \textit{CC chondrites---CL, CV, CO, CM, CR, CI, Tagish Lake (TL).} We calculated the proportions of end-members in CC chondrites using their bulk $\varepsilon^{54}$Cr and $\varepsilon^{50}$Ti values \citep{Dauphas:2024p115805,Petitat:2011p23,Trinquier:2009p324}. In these cases, $\varepsilon^{50}$Ti primarily reflects \(P_{\mm{CAI}}\) and $\varepsilon^{54}$Cr predominantly reflects \(P_{\mm{NC}}\) and \(P_{\mm{CI}}\). 
    \item \textit{NC iron meteorites---IC, IIAB, IIIAB, IIIE, IVA, ungrouped.} NC chondrites contain negligible amounts of CAIs ($\sim$0.01 area\%\ ; \citealt{Dunham:2023p643}). As such, we assume \(P_{\mm{CAI}}=0\) for all NC meteorites, enabling us to calculate \(P_{\mm{NC}}\) and \(P_{\mm{CI}}\) from a single isotopic signature. For NC iron meteorites, the relatively large uncertainties on their $\varepsilon^{62}$Ni values makes it difficult to recover accurate values of \(P_{\mm{NC}}\) and \(P_{\mm{CI}}\) following the same approach as CC iron meteorites. Instead, we adopted the $\varepsilon^{54}$Cr values of chromite and daubréelite grains extracted from NC magmatic iron meteorites \citep{Anand:2021p6}. Although these concentrations have only been measured from three of the five NC magmatic iron meteorite groups, their lower uncertainties permitted more reliable values of \(P_{\mm{NC}}\) and \(P_{\mm{CI}}\) to be calculated. Throughout this study, we chose not to consider NC ungrouped iron meteorites whose apparent differentiation ages were so young that they likely reflect mixing during impacts \citep{Spitzer:2025P134}.
    \item \textit{NC chondrites---H, L, LL, R, EH, EL.} We used bulk $\varepsilon^{54}$Cr concentrations to calculate \(P_{\mm{NC}}\) and \(P_{\mm{CI}}\) in NC chondrites.
\end{itemize}
See the Supplementary Information for a discussion on the reasons we chose these elements and how our results could be mapped onto elements that exhibit different trends. 


\subsection{Thermal evolution model}
We use an analytical model to calculate planetesimal accretion times from the PMTs of chondrites and differentiation ages of iron meteorites. The complete mathematical description, approximations, and assumptions are discussed in the Supplementary Information.

This model calculates the time taken for a planetesimal to heat via the radioactive decay of $^{26}$Al from its accretion temperature to either its peak temperature or differentiation temperature in seven stages: 

\begin{enumerate}
    \item From the initial temperature to the temperature of ice melting; \(T_{\mm{init}}\) - 272 K, where \(T_{\mm{init}}\) is the initial temperature of the accreting material.
    \item During ice melting; 272 - 273 K.
    \item During silicate hydration; 273 K - \(T_{\mm{hyd}}\), where \(T_{\mm{hyd}}\) is the temperature at which silicate hydration ends.
    \item During silicate dehydration; \(T_{\mm{hyd}}\) - 1223 K.
    \item From the end of silicate dehydration to the start of Fe--FeS melting; 1223 - 1260 K.
    \item During Fe--FeS melting; 1260 - 1400 K.
    \item During Fe--FeS and silicate melting, to the temperature of 30\%\ silicate melting (see Supplementary Information); 1400 - 1520 K.
\end{enumerate}
Once the body reaches 30\%\ silicate melting, we argue that Fe--FeS rain-out will have occurred, so the body will have differentiated \citep{Sanderson:2024p116323}. 

The temperature change due to the decay of $^{26}$Al will have depended on both the concentration of Al (\(X_{\mm{Al}}\)) and the specific heat capacity (\(C_p\)) of a planetesimal \citep{Ghosh:1999p121,Merk:2002p183}. Because planetesimals experienced several phase changes and chemical reactions as they heated, the expressions for \(X_{\mm{Al}}\) and \(C_p\) will have varied between the different stages of our model (see Supplementary Information). Using these varying expressions, we calculated values of \(C_p\) and \(X_{\mm{Al}}\) in each relevant stage of the model for planetesimals built from each meteorite group. We then used \(C_p\) and \(X_{\mm{Al}}\) to calculate the energy change per unit mass, \(\Delta E=\int_{T_{\mm{lower}}}^{T_{\mm{upper}}}\frac{C_p}{X_{\mm{Al}}}dT\) in each relevant stage using numerical integration. \(T_{\mm{lower}}\) is the lower temperature bound on each stage of the model. \(T_{\mm{upper}}\) is either the upper bound on a stage of the model or the PMT reached by a meteorite group. Finally, we calculated the accretion time of the planetesimal from the time taken for the total energy change per unit mass (i.e., the sum of these integrals over all relevant stages of the model; see Supplementary Information) to be produced by the decay of $^{26}$Al. The differentiation ages and PMTs we adopt are included in Tables S3 and S4 and are discussed further in the Supplementary Information.

Before differentiating, the parent planetesimals of iron meteorites and rocky achondrites are believed to have been chondritic, i.e., composed of chondrules, matrix, and CAIs. Combined with the natural variations in the fractions of these components among different chondrite groups \citep{Scott:2014TreatiseGeoChem}, we expect the bulk values of \(C_p\) and \(X_{\mm{Al}}\) will have differed among planetesimals built from different meteorites. These fractions can be recovered from chondrites via petrographic observations. However, these fractions cannot be gleaned from such observations of iron meteorites (or rocky achondrites) because their pre-existing chondritic textures were lost when their parent planetesimals melted. As such, we adopted different approaches to recover the fractions of components within a parent planetesimal:
\begin{itemize}
    \item \textit{CC iron meteorites.} As introduced in section \ref{mix}, we calculated the fractions of chondrules, matrix, and CAIs in the chondritic progenitor to CC iron meteorites from their bulk $\varepsilon^{62}$Ni and $\Delta^{95}$Mo values.
    \item \textit{CC chondrites.} To be consistent with CC iron meteorites, we also calculated the fractions of chondrules, matrix, and CAIs in CC chondrites from their bulk isotopic compositions. In these cases, we used $\varepsilon^{54}$Cr and $\varepsilon^{50}$Ti values.
    \item \textit{NC iron meteorites.} Unlike CC meteorites, the near-identical isotopic compositions of NC bulk chondrites and chondrules \citep{Schneider:2020p116585,Kita:2010p6610} demonstrate that matrix in these meteorites shares the same isotopic composition as chondrules. As such, it is not possible to distinguish the proportion of chondrules and matrix in the progenitors of NC magmatic iron meteorites from their bulk isotopic compositions. Instead, we calculated two sets of accretion ages for the parent bodies of these meteorites: one with 5 wt\%\ matrix; and one with 95 wt\%\ matrix. The former is similar to the observed matrix mass fraction in ordinary chondrites (OCs; \citealt{Bryson:2021p163}), so is more likely to be realistic; the latter provided a reasonable early limit on accretion times. 
    \item \textit{NC chondrites.} For the same reasons as NC iron meteorites, we could not calculate the fractions of chondrules and matrix in NC chondrites. Instead, we adopted their measured Al concentrations alongside their proportions of chondrules and matrix derived from petrographic observations.
\end{itemize}

We were not able to calculate precise formation ages for rocky achondrite or primitive achondrite parent planetesimals. This is due to the absence of a geochronometer among these meteorites that yields definitive dates of an event/process from which formation ages could be calculated (e.g., \citealt{Anand:2023p126004}). However, the (partially) melted and differentiated natures of these meteorites means the accretion ages of their parent planetesimals will have been similar to iron meteorites from the same reservoir. As such, we adopt the same, broad formation time for all rocky achondrite parent bodies from the same reservoir, which matches to the range of values we recover from corresponding magmatic iron meteorites.

Our thermal model contains sources of error that could impact the accretion ages we calculate. These are all discussed in detail in the Supplementary Information. The dominant sources are the analytical uncertainties on the differentiation ages of iron meteorites, the PMTs of chondrites, and the nucleosynthetic isotope anomalies of both types of meteorites. The error bars we present in our figures are calculated by taking extreme combinations of these parameters and calculating accretion ages. Other sources of error are likely smaller these error bars, so introduce uncertainties that fall within these ranges. Additionally, the values of $P_{CI}$, $P_{CAI}$, and $P_{NC}$ we calculate are used as part of the calculation of the fractions of chondrite components in our accretion age model. These shared inputs could introduce a correlation between $t_a$ and $P_{CI}$, $P_{CAI}$, and $P_{NC}$. Crucially, however, our $t_a$ calculation includes an independent parameter: differentiation times for iron meteorites; PMTs for chondrite parent bodies. As such, different sets of inputs are used to calculate $P_{CI}$, $P_{CAI}$, and $P_{NC}$ compared to $t_a$, so any correlation is not a direct consequence of the calculations sharing identical inputs.

\section{Results}
\label{Results}

\subsection{Accretion Ages of Iron Meteorite Parent Bodies}
\label{IronMetRes_ta}

\begin{figure}[t]
\centering
\includegraphics[scale=0.95]{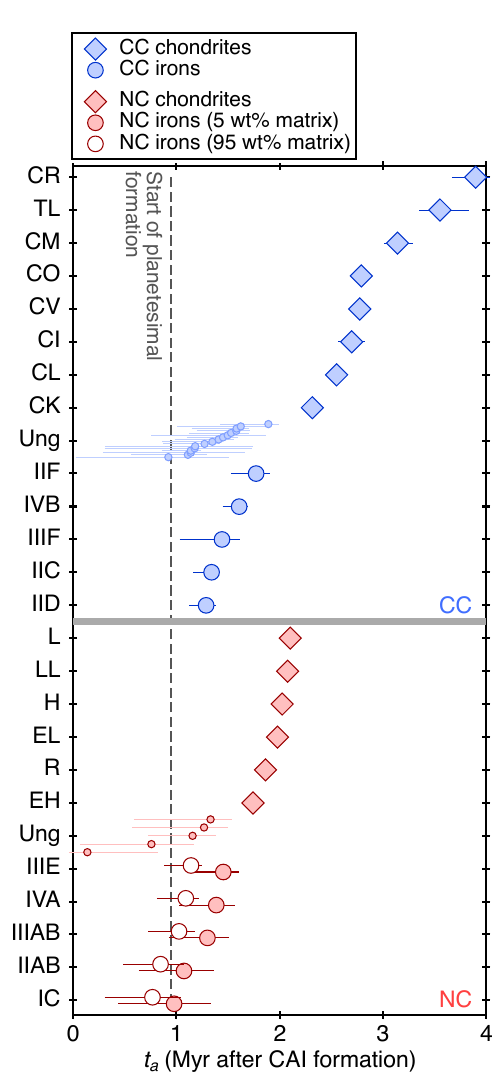}
\caption{Calculated accretion ages (\(t_\mm{a})\) for NC (red) and CC (blue) meteorites. The error bars are smaller than the points in some cases. The grouped iron meteorites and chondrites are ordered by their accretion age within each reservoir. The ungrouped iron meteorites are ordered by their accretion ages within each reservoir and are placed between the grouped iron meteorites and grouped chondrites in their respective reservoir. Ages calculated for NC iron meteorites with 95\%\ matrix (closed symbols) and 5\%\ matrix (open symbols) in their chondritic progenitor are included. Our adopted start time of planetesimal formation is marked by the vertical dashed line. The grey line distinguishes NC and CC meteorites.} \label{Ta}
\end{figure}

Our calculated accretion ages of iron meteorite parent planetesimals are included in Tables \ref{GroupedOutputs} and \ref{UngroupedOutputs} and are shown in Fig. \ref{Ta}. Among CC iron meteorites, the range of accretion ages recovered from the magmatic groups ($\sim$1.29 - 1.77 Myr after CAI formation) is similar to that of ungrouped meteorites ($\sim$0.92 - 1.89 Myr after CAI formation). The CC ungrouped iron meteorites include one notably old member (ILD 83500) and one notably young member (Tucson). However, the uncertainties on the recovered ages of these two meteorites overlap appreciably with the ages recovered from the other ungrouped CC iron meteorites, so are unlikely to be anomalous.

Among NC magmatic iron meteorites, the range in possible matrix proportions in their progenitors causes the accretion ages of their parent planetesimals to vary by 0.20 - 0.31 Myr depending on the group. This change does not affect the relative order of the ages of NC magmatic iron meteorite groups. As such, the amount of matrix in the progenitor to the NC iron meteorites has a secondary effect on their formation ages, so we adopt 5 wt\%\ matrix because this is similar to the amount observed in NC chondrites. Among the NC ungrouped iron meteorites, EET 83230 has a very old calculated accretion age, which likely reflects mixing during impacts (similar to the sHH and sHL subgroups of the IAB iron meteorites; \citealt{Spitzer:2025P134}). Another example, Zacatecas (1792), is also notably old, but is not as clearly anomalous. As such, it is uncertain whether this value is a true age or the result of impacts. The calculated accretion ages of the remaining NC ungrouped iron meteorites occupy a similar range ($\sim$1.16 - 1.33 Myr after CAI formation) to the grouped NC magmatic iron meteorites ($\sim$0.98 - 1.45 Myr after CAI formation). We assign the formation age of all NC rocky achondrites as this latter range.

Unlike previous models \citep{Kruijer:2017p6712}, we find that the accretion ages of NC iron meteorite and CC iron meteorite parent bodies occupy similar, overlapping ranges. Additionally, the earliest accretion ages of iron meteorite parent bodies in both reservoirs are similar ($\sim$0.95 Myr after CAI formation), which is notably younger than previous models. These differences are largely due to updated differentiation ages of iron meteorites \citep{Spitzer:2021p117211,Hellmann:2024p118518,Spitzer:2025P134} as well as variations in \(C_p\) and \(X_{\mm{Al}}\) among the progenitors of iron meteorites introduced by our model.

\subsection{Accretion Ages of Chondrite Parent Bodies}
\label{ChdRes_ta}

Our formation ages of chondrite parent bodies calculated from their PMTs are included in Fig. \ref{Ta} and Table \ref{GroupedOutputs}. For NC chondrites, these ages fall between $\sim$1.74 - 2.10 Myr after CAI formation. For CC chondrites, they fall between $\sim$2.32 - 3.90 Myr after CAI formation. These formation age windows do not overlap, although this window is narrow for NC chondrites. The older age range of NC chondrites is due to their paucity of CAIs and higher PMTs than most CC chondrites. This paucity of CAIs reduces the $^{26}$Al concentration in NC meteorites, requiring earlier accretion ages to reach these higher PMTs.

\subsection{Proportions of CI material}
\label{ChdRes_PCI}

\begin{figure}[t]
\centering
\includegraphics[scale=1]{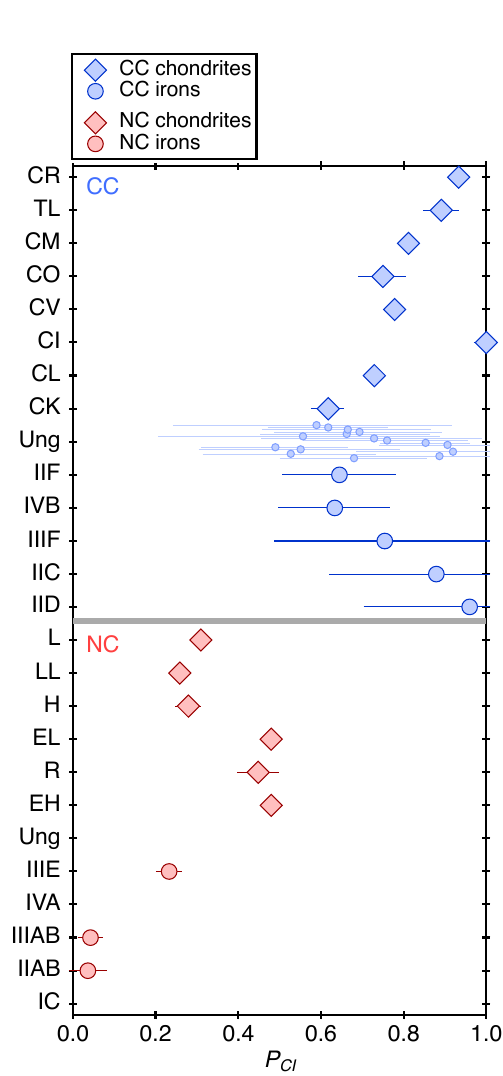}
\caption{Recovered proportions of CI material (\(P_{\mm{CI}}\)) for NC (red) and CC (blue) meteorites. The error bars are smaller than the points in some cases. The order of meteorites is the same as Fig. \ref{Ta}. The grey line distinguishes NC and CC meteorites.}\label{P_CI}
\end{figure}

Our nominal values of $P_{\mm{CI}}$ occupy non-overlapping ranges between NC and CC meteorites (Fig. \ref{P_CI} and Tables \ref{GroupedOutputs} and \ref{UngroupedOutputs}). CC meteorites contain higher proportions of CI material ($\sim$0.49 - 1.00) than NC meteorites ($\sim$0.04 - 0.48). Among CC meteorites, iron meteorites and chondrites occupy overlapping ranges. Among NC meteorites, iron meteorites and chondrites occupy non-overlapping ranges. The recovered values for CC ungrouped iron meteorites (0.49 - 0.92) occupy a similar range to the CC grouped iron meteorites (0.63 - 0.96).

\subsection{Proportions of CAI material}
\label{ChdRes_PCAI}

\begin{figure}[t]
\centering
\includegraphics[scale=1]{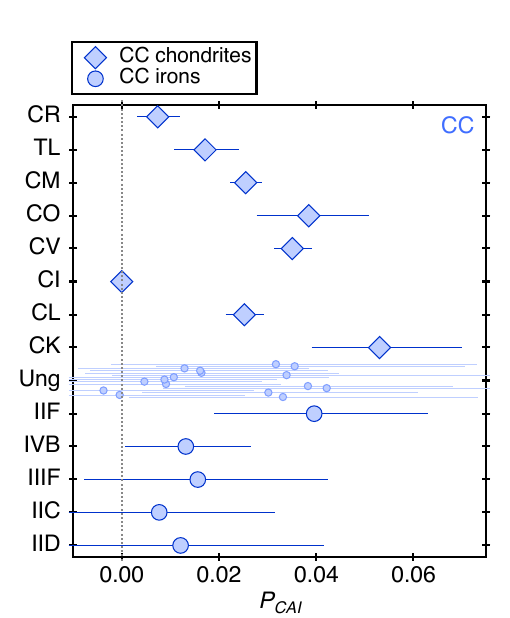}
\caption{Recovered proportions of CAI material (\(P_{\mm{CAI}}\)) for CC meteorites. The error bars are smaller than the points in some cases. The order of meteorites is the same as Fig. \ref{Ta}.}\label{P_CAI}
\end{figure}

Among CC meteorites, our recovered values of $P_{\mm{CAI}}$ range from -0.004 - 0.053 (Fig. \ref{P_CAI} and Tables \ref{GroupedOutputs} and \ref{UngroupedOutputs}). The nominal negative values recovered for two CC ungrouped iron meteorites reflect their low $\Delta^{95}$Mo values. However, the uncertainties on these values are large enough that, in reality, these meteorites likely have small positive values of $P_{\mm{CAI}}$. The values we calculate for each group are similar to the volume percentages of these inclusions recovered from petrographic observations of CC chondrites \citep{Scott:2014TreatiseGeoChem,Dunham:2023p643}. CC magmatic iron meteorites and chondrites occupy overlapping ranges of $P_{\mm{CAI}}$. The range of values recovered from CC ungrouped iron meteorites (-0.004 - 0.042) is marginally larger than that recovered for the CC grouped iron meteorites (0.008 - 0.040).

\section{Discussion}
\label{Discussion}

The accretion times, proportions of CI material, and proportions of CAI material vary among meteorite parent planetesimals in the NC and CC reservoirs. These reservoirs are characterised by pronounced differences in nucleosynthetic isotopic compositions \citep{Warren:2011p93,Burkhardt:2021peabj7601}. These differences argue that distinct regions in our protoplanetary disk were isolated from each other for at least part of the disk's lifetime. This division requires a barrier in the disk during this time window \citep{Kruijer:2017p6712} that prevented the creation of an homogenised, disk-wide isotopic ratio for each element. Based on the water concentrations of primitive chondrites \citep{Alexander:2018p36,Vacher:2020p53}, the NC reservoir was likely closer to the Sun and the CC reservoir was likely further from the Sun. Material at the furthest heliocentric distances within the CC reservoir is believed to have had isotopic compositions that match CI chondrites (e.g., \citealt{Bryson:2021p163,Yap:2023p115680}).

\subsection{CC Meteorites: $P_{\mm{CI}}$}
\label{CCDiss}

Collectively, the parent bodies of CC meteorites occupy the wider range of accretion ages, starting at $\sim$0.92 Myr after CAI formation and ending at $\sim$3.90 Myr after CAI formation. This longevity argues that gas and dust were long-lived in the outer disk, consistent with the small grain size (<1 - 80 µm) in some chondrules in CR chondrites \citep{Schrader:2018p30}.

\begin{figure}[t]
\centering
\includegraphics[scale=0.9]{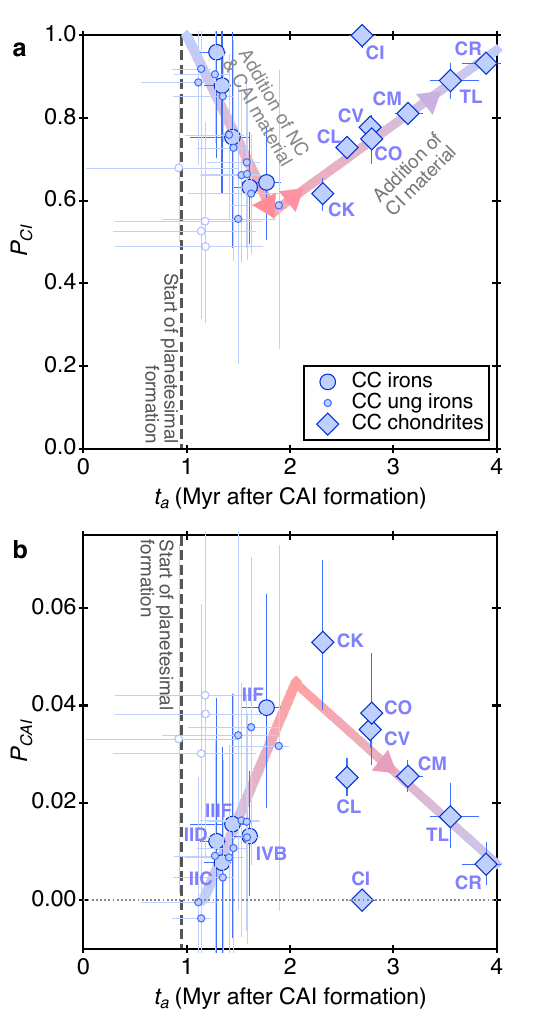}
\caption{a \(P_{\mm{CI}}\) as a function of \(t_{\mm{a}}\) for CC meteorites. The trends we identify are highlighted by arrows that evolve from blue to red as the amount of NC and CAI material in the CC reservoir increases. The names of CC iron meteorites are not included to avoid overcrowding the figure. b \(P_{\mm{CAI}}\) as a function of \(t_{\mm{a}}\) for CC meteorites. The error bars are smaller than the size of the points in some cases. Our recovered start time of planetesimal formation is marked by the vertical dashed line. The lines are linear best fits to iron meteorite (excluding the four ungrouped examples that did not fall on the main trend) and chondrite (excluding CI chondrites) data points. The ungrouped iron meteorite data points that as discussed in the text as falling off the main trend are shown as open symbols.}\label{P_v_ta_CC}
\end{figure}

By combining our values of $t_{\mm{a}}$ with our values of \(P_{\mm{CI}}\) and \(P_{\mm{CAI}}\), we can explore the evolution of the abundances of CI and CAI material throughout the disk. Fig. \ref{P_v_ta_CC}a shows the evolution in $P_{\mm{CI}}$ as a function of $t_{\mm{a}}$ in the CC reservoir. This variation includes an initial steep decrease in the amount of CI material across $\sim$0.8 Myr observed in both CC grouped iron meteorites and the majority of CC ungrouped iron meteorites. This decrease is followed by a shallower recovery over the next $\sim$2 Myr captured by CC chondrites. These trends are refined compared to those of \cite{Bryson:2021p163}, with more reliable values of $P_{\mm{CI}}$ and $t_{\mm{a}}$ as well as the inclusion of ungrouped iron meteorites.

\begin{figure}[t]
\centering
\includegraphics[scale=0.85]{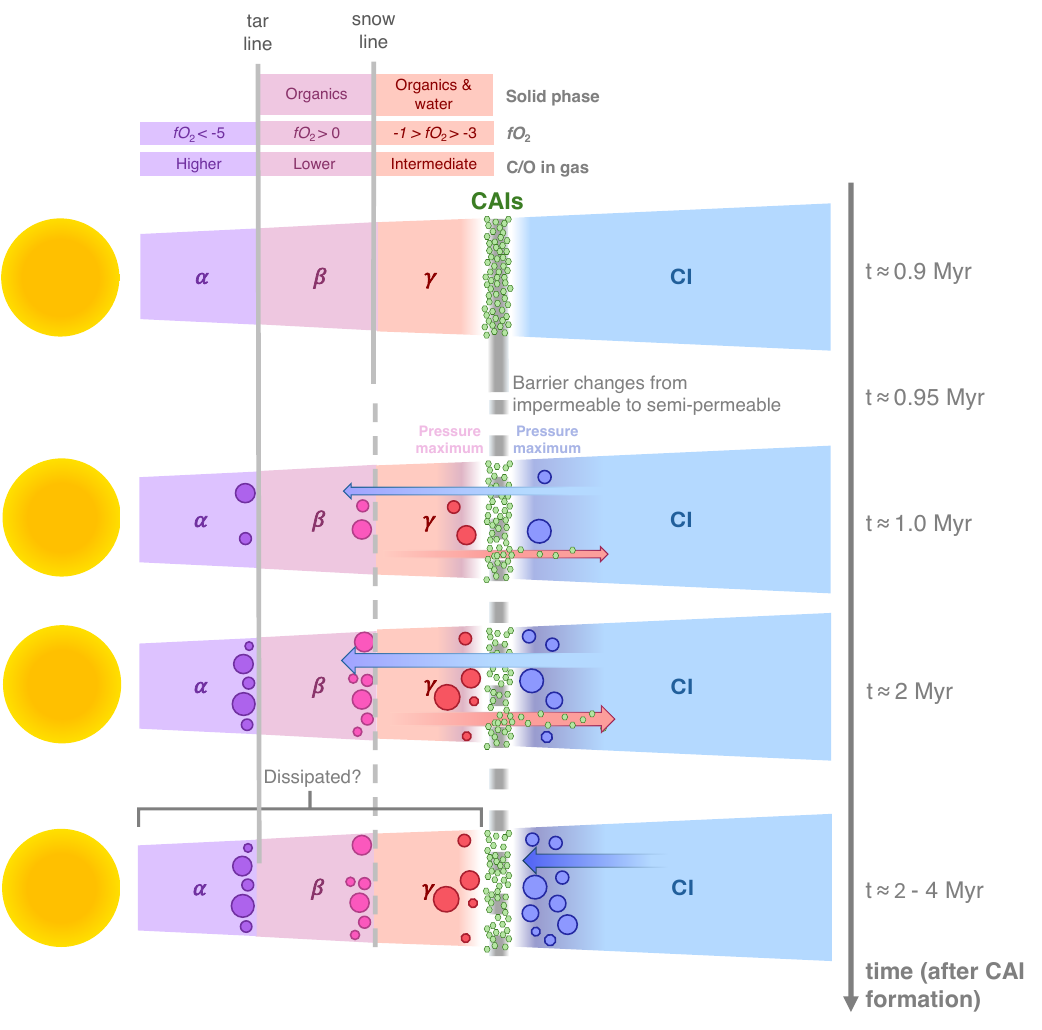}
\caption{\scriptsize{Schematic of disk evolution and planetesimal formation. Timings are Myr after CAI formation. At early times ($\sim$0.9 Myr after CAI formation) the disk is divided by a barrier feature that creates an inner reservoir that is divided into three sub-reservoirs---NC-$\alpha$, NC-$\beta$, and NC-$\gamma$---and an outer reservoir composed of CI material. CAIs are localised at the position of this barrier. At $\sim$0.95 Myr after CAI formation, the nature of the barrier changes, becoming semi-permeable. After this time (labelled nominally as $\sim$1.0 Myr after CAI formation), this feature has three affects: it transports NC material outwards and CC material inwards (shown by the arrows); it causes planetesimals to form on both sides (shown by circles); and it moves CAIs overwhelmingly outwards (shown by the green hexagons). The locations of planetesimals within the NC-$\alpha$ and NC-$\beta$ sub-reservoirs are arbitrary. By $\sim$2 Myr after CAI formation, more NC and CI material has crossed the feature and more planetesimals have formed. Between $\sim$2-4 Myr after CAI formation, NC material no longer mixes into the CC reservoir, and distal CI material starts mixing into the innermost CC reservoir. More planetesimals form during this period in the CC reservoir. The NC reservoir may have dissipated by this time. Throughout this figure, the feature that separates the NC and CC reservoirs is depicted by a thick grey line that is solid when impermeable and dashed when semi-permeable. The condensation lines within the NC reservoir are shown by thin grey lines that are solid when impermeable and dashed when semi-permeable.}}\label{schem}
\end{figure}

These trends provide several insights into the motion of gas, dust, CAIs, and chondrules in our disk. To facilitate discussion, Fig. \ref{schem} illustrates how the isotopic evolutions we infer could have physically arisen. Firstly, shortly before $\sim$0.95 Myr after CAI formation, the CC reservoir appears to have been composed almost entirely of CI-like material (Fig. \ref{P_v_ta_CC}). Because this early signature predates the formation of planetesimals, we cannot use the trends we recover to illuminate its origin; this will instead be the topic of a future study. Next, the isotopic composition of the CC reservoir started to evolve, becoming more NC-like over a period of $\sim$0.8 Myr. The most straightforward mechanism of achieving this change is for NC material to mix into the CC reservoir. This mixing is supported by the isotopic compositions of individual chondrules from CC chondrites, which demonstrate that (parts of) these spherules can exhibit NC isotopic compositions \citep{VanKooten:2021p70,Schrader:2020p133,Williams:2020p05235}. Finally, the recovery to more CI-like isotopic compositions can be explained by mixing of CI material from the distal disk that dominates the earlier flux from the NC reservoir (Fig. \ref{schem}). 

This isotopic evolution is most readily explained by material travelling across the barrier in our disk, i.e., the barrier was at least semi-permeable (Fig. \ref{schem}). However, transport and mixing between the NC and CC reservoirs is inconsistent with the pronounced difference in isotopic composition between these reservoirs at $\sim$0.95 Myr after CAI formation. These two behaviours can be reconciled by the nature of the barrier changing at $\sim$0.95 Myr after CAI formation, from impermeable to at least semi-permeable. Within this framework, the sizeable difference in isotopic compositions between the reservoirs was introduced between $\sim$0 - 0.95 Myr after CAI formation and the isotopic variation within each reservoir results from motion across the semi-permeable barrier at $\gtrsim$0.95 Myr after CAI formation. While the feature that served as this (semi-permeable) barrier is currently unknown, several candidates have been proposed, including Jupiter \citep{Kruijer:2017p6712}, the snow line \citep{Lichtenberg:2021p365}, or the disk's expansion \citep{Liu:2022pealm3045}. The isotopic evolutions and behaviours we have identified here can now be utilised as novel constraints to help models distinguish between these candidates.

Four CC ungrouped iron meteorites deviate from the main $P_{\mm{CI}}$ trend: Babb's Mill (Troost's Iron), ILD 83500, Piñon, and Grand Rapids; the first two are part of the South Byron trio \citep{Mccoy:2019p358}. All four of these meteorites contain lower proportions of CI material (i.e., higher proportions of NC material) than other CC iron meteorites with similar accretion ages, as well as exhibit anomalously early differentiation ages \citep{Spitzer:2025P134}. It is plausible that these four meteorites could be part of a second trend that is steeper than the main trend. However, the low number of meteorites in this grouping makes this trend's existence difficult to determine with certainty. If it does exist, this trend would argue these meteorites contain higher proportions of material that derive from smaller heliocentric distances. This behaviour is most readily achieved if these meteorites originate from closer to the semi-permeable barrier than other CC meteorites. As such, there may be multiple sources of CC meteorites: one slightly closer to the semi-permeable barrier that generates a minority of CC planetesimals; and one slightly further from the semi-permeable barrier that generates the majority of CC planetesimals.

Additionally, CI chondrites themselves fall off the main $P_{\mm{CI}}$ trend, arguing they do not originate from the same location as other CC meteorites. The outer portion of the CC reservoir is believed to have been composed of dust with pure CI isotopic compositions (\citealt{Bryson:2021p163}). As such, the CI chondrites likely originate from the distal disk, representing a final source of CC meteorites. Given their isotopic similarities to CI chondrites, it is possible that asteroids Ryugu \citep{Yokoyama:2022pabn7850} and Bennu \citep{Lauretta:2024p2453,Barnes:2025aa} also originate from this distal source.

\subsection{CC Meteorites: $P_{\mm{CAI}}$}
\label{CAIDiss}

Our model also yields the evolution of $P_{\mm{CAI}}$ in the CC reservoir (Fig. \ref{P_v_ta_CC}b). This includes an initial steep increase over $\sim$0.9 Myr, followed by a later, shallow decrease over $\sim$1.9 Myr. The steep increase is recorded by CC grouped iron meteorites as well as most CC ungrouped iron meteorites. As such, across both CC iron meteorites and chondrites, $P_{\mm{CAI}}$ negatively correlates with $P_{\mm{CI}}$ (Figs. \ref{P_v_ta_CC}, S4). This behaviour indicates that CAIs mixed into the CC reservoir alongside NC material and were also diluted by the addition of CI material at later times. Interestingly, however, NC chondrites contain orders-of-magnitude lower abundances of CAIs than CC meteorites \citep{Dunham:2023p643}. This paucity implies that CAIs were a negligible component of the material that left the NC reservoir when it started its journey into the CC reservoir. As such, CAIs must have been acquired in transit as the NC material travelled into the CC reservoir. This behaviour argues that CAIs resided at the radial distance of the barrier between the NC and CC reservoirs at $\sim$0.95 Myr after CAI formation (Fig. \ref{schem}). Their refractory mineralogy/composition and exotic isotopic signatures demonstrate that CAIs formed comparatively close to the Sun \citep{Ebel:2000p339,Bekaert:2021peabg8329}. As such, these refractory inclusions appeared to have travelled from near the Sun to the radial distance of the barrier within $\sim$0.95 Myr of their formation where they became localised. The reason the barrier is able to concentrate and confine CAIs during their first $\sim$0.95 Myr is unknown. However, the barrier's influence on the distribution of CAIs provides another novel constraint on its earliest behaviour that can aid in illuminating its source. CAIs then left this radial distance at the same time that NC material started mixing into the CC reservoir. As such, the change in the permeability of the barrier at $\sim$0.95 Myr after CAI formation coincided with a change in this feature's ability to confine and store CAIs.

The same four CC ungrouped iron meteorites that fall off the main \(P_{\mm{CI}}\) trend also fall off the main \(P_{\mm{CAI}}\) trend (Fig. \ref{P_v_ta_CC}b). These meteorites contain higher proportions of CAIs than other CC iron meteorites with similar values of $t_a$. Because CAIs seemingly mix into the CC reservoir from the location of the semi-permeable barrier, this behaviour is consistent with our proposal that this subset of meteorites originate from closer to this feature than other CC meteorites.

Equally, CI chondrites fall off the main $P_{\mm{CAI}}$ trend (Fig. \ref{P_v_ta_CC}b), containing effectively 0 wt$\%$ CAIs. This very low abundance agrees with petrographic observations of these meteorites \citep{Frank:2023p1495}. Because CAIs appear to enter the CC reservoir from its inner-most edge, the paucity of these inclusions in CI chondrites supports our proposal that these meteorites originate from the distal disk.

\subsection{NC Meteorites}
\label{NCDiss}

The parent bodies of NC meteorites occupy the narrower and earlier range of $t_a$, starting at $\sim$0.98 Myr after CAI formation and ending at $\sim$2.10 Myr after CAI formation. These early times imply that planetesimal formation in the NC reservoir was temporally limited, which is consistent with the NC reservoir dissipating before the CC reservoir. Disk dissipation can result from several process, including the accretion of disk material onto the Sun and/or ejection of material by magnetised winds \citep{Alexander:2014p475}. This early dissipation of the NC reservoir would coincide with the change in slope in the evolutions of $P_{\mm{CI}}$ and $P_{\mm{CAI}}$ among CC meteorites (Fig. \ref{P_v_ta_CC}). Indeed, this change is expected if the NC reservoir no longer existed to mix into the CC reservoir. The clearing of inner regions of disks before outer regions has also been observed in distant disks with the ALMA telescope \citep{Curone:2025pL9}. Hence, the processes that drove dissipation in these disks may have been similar to those that acted in our disk. 

\begin{figure}[t]
\centering
\includegraphics[scale=1.0]{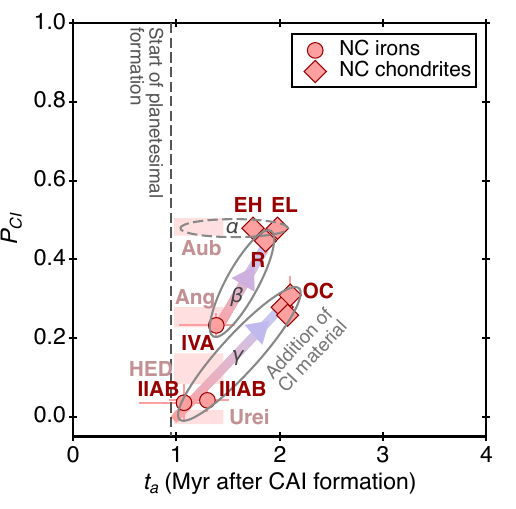}
\caption{\(P_{\mm{CI}}\) as a function of \(t_{\mm{a}}\) among NC meteorites. The data ranges of the different sub-reservoirs (NC-$\alpha$, NC-$\beta$, and NC-$\gamma$) are depicted by ellipses. The ellipse for the NC-$\alpha$ sub-reservoir is dashed because its position and shape are not constrained by an iron meteorite group. The trends we identify are included as arrows that evolve from red to blue as the amount of CI material in the NC reservoir increases. The NC rocky achondrites are shown as light red rectangles. The extent of these rectangles in \(t_{\mm{a}}\) matches our calculated range of \(t_{\mm{a}}\) for NC magmatic iron meteorites. The vertical extent of these rectangles corresponds to the errors on the calculated values of $P_{\mm{CI}}$ for each NC rocky achondrite group (Table S6). The error bars are smaller than the size of the points in some cases. Our recovered start time of planetesimal formation is marked by the vertical dashed line.}\label{P_v_ta_NC}
\end{figure}

Unlike CC meteorites, the values of \(P_{\mm{CI}}\) among NC meteorites collectively do not form a clear, single trend (Fig. \ref{P_v_ta_NC}). If these proportions are interpreted as a single trend, it would include particularly large variations over very short times (e.g., from the IIIAB to IVA iron meteorites, from ureilites to aubrites, and from EL to H chondrites). Instead, the NC reservoir has been proposed to have contained sub-reservoirs \citep{Rufenacht:2023p110}, arguing that at least part of the isotopic variation among NC meteorites is spatial in origin. Moreover, contrary to CC chondrites, NC chondrites exhibit a particularly wide range of redox states, with their inferred oxygen fugacities ($fO_{2}$) varying by a factor of $\sim10^8$ \citep{Righter:2016p1928}. Broadly, NC meteorites fall into three groupings in terms of $fO_{2}$: 
\begin{itemize}
    \item similar to enstatite chondrites (ECs); $fO_{2}$<-5 relative to the iron-wüstite (IW) buffer; accompanying rocky achondrite groups are very reduced, e.g., aubrites.
    \item similar to OCs; -1>$fO_{2}$>-3 relative to the IW buffer; accompanying achondrite groups contain metal, e.g., HEDs, ureilites, and acapulcoites and lodranites. 
    \item similar to Rumuruti chondrites (RCs); $fO_{2}$>0 relative to the IW buffer; accompanying achondrite groups contain metal oxides, e.g., angrites and possibly brachinites.
\end{itemize}
Among the most pristine chondrites in these groupings, these redox states can be ascribed to conditions in the disk, arguing that the NC reservoir contained several regions with pronounced differences in redox state. Following \cite{Bermingham:2020p133}, these redox conditions could result from condensation lines within the NC reservoir, namely the tar line and ice line. Specifically, sunward of the tar line, organic molecules and water will have both existed as a gas (Fig. \ref{schem}). This enrichment in carbon would have made this gas comparatively reducing. Chondrules that interacted with this gas when molten are therefore expected to exhibit reduced mineralogies. This is consistent with the presence of niningerite, alabandite, oldhamite, and Si-bearing kamacite in ECs \citep{Weisberg:2012p101}. In the region between the tar and ice lines, water will have still existed as a gas while appreciable amounts of carbon will have been locked in solid phases. As such, the gas in this region will have been enriched in oxygen relative to carbon, meaning that chondrules in this region will have interacted with a more oxidised gas. This will have led to an oxidised mineralogy in these spherules, consistent with the elevated Fe contents of silicates in primitive RCs \citep{Bischoff:2011p101}. Moreover, the absence of ice in this region means the RC parent body is expected to have contained very little water. This is supported by the low abundances of phases in pristine clasts in RCs that are clear indicators of aqueous alteration in other chondrites, e.g., magnetite, carbonates, and phyllosilicates \citep{Bischoff:2011p101}. Outside the snow line, both water and organic molecules will have been solid, which could have resulted in a gas with an intermediate redox state, consistent with OCs. In this scenario, the OC parent bodies would have accreted ice, which is supported by the presence of magnetite, carbonates, and phyllosilicates in primitive examples of these meteorites \citep{Krot:1997p219}. As observed in carbonaceous chondrites \citep{Rubin:2019p125528}, these phases can be lost or absent due to heating on the parent body, potentially explaining why they are not widespread among OCs.

Dividing NC meteorites into these groupings, which we term $\alpha$ (EC-like $fO_{2}$), $\beta$ (RC-like $fO_{2}$), and $\gamma$ (OC-like $fO_{2}$), produces clearer trends in the isotopic evolution within the NC reservoir (Fig. \ref{P_v_ta_NC}). Firstly, the NC-$\alpha$ sub-reservoir contains the most CI-like initial isotopic composition and does not exhibit a clear isotopic evolution, as shown by aubrites and ECs. No iron meteorites are known to originate from this sub-reservoir. Secondly, the NC-$\beta$ sub-reservoir shows a less CI-like initial composition and a small evolution towards CI material over time, as evidenced by angrites and RCs. The IVA iron meteorites appear to fall on this trend, suggesting they may be an iron meteorite group that originates from the NC-$\beta$ sub-reservoir. This association is nominally supported by the high average Ni and Co concentrations of IVA iron meteorites, which suggest they are among the more oxidised NC iron meteorites \citep{Rubin:2018p2357,Spitzer:2025P134,Grewal:2024p290}. Finally, the NC-$\gamma$ sub-reservoir initially has the lowest proportion of CI material and exhibits a relatively large evolution towards CI material over time. This is demonstrated by the difference in isotopic composition between ureilites, HEDs, and OCs. The IIAB and IIIAB iron meteorites fall on this trend, suggesting they could originate from this sub-reservoir. The origin of the initial signatures within each sub-reservoir will be the topic of a future study.

Similar to CC meteorites, these isotopic evolutions among NC meteorites can be explained by the motion of material across the barrier in the disk (Fig. \ref{schem}). In this case, this transport involved the inward addition of CC material into the NC reservoir. Specifically, the largest amount of CC material was added to the NC-$\gamma$ sub-reservoir, a smaller amount was added to the NC-$\beta$ sub-reservoir, and a minimal amount (potentially none) was added to the NC-$\alpha$ sub-reservoir. Within this framework, condensation lines appear to hinder the passage of material through the NC reservoir \citep{Drazkowska:2017pA92}, having a greater effect further from the CC reservoir. Additional evidence for this transport has recently been identified in OCs and RCs, where smaller chondrules tend to exhibit isotopic compositions that are closer to CC material than larger chondrules \citep{Marrocchi:2024p52}. This observation implies a size-dependence to inward transport, consistent with the small sizes of the rare CAIs in NC chondrites \citep{Dunham:2023p643}. In contrast to CC meteorites, these isotopic changes in the NC reservoir simply evolve in one direction, towards CI material over time. Combined with the evolution among CC meteorites, the barrier in our disk appears to have become semi-permeable in both directions at very similar times ($\sim$0.95 Myr after CAI formation). As such, the trends among NC meteorite provide unique, additional constraints that models of disk evolution can now utilise to help identify the feature that acted as the semi-permeable barrier.

\subsection{The Trigger for Planetesimal Formation}
\label{TiggerDiss}

The isotopic evolutions we identify among both NC and CC meteorites start at similar times, which we nominally take as $\sim$0.95 Myr after CAI formation (Figs. \ref{P_v_ta_CC} and \ref{P_v_ta_NC}). This similarity argues that NC material started mixing into the CC reservoir at a very similar time that CC material started mixing into the NC reservoir. This time is also coincident with the start of planetesimal formation in both reservoirs (Fig. \ref{Ta}). These coeval timings argue that the feature that drove motion between the NC and CC reservoirs also kickstarted planetesimal formation in both reservoirs. Hence, any phenomenon that is found to recreate the isotopic evolutions we identify can also be regarded as the trigger for planetesimal formation. The conditions within pressure maxima are conducive to planetesimal formation \citep{Morbidelli:2024pA147}. Therefore, the ability of this feature to create pressure maxima could be pivotal to the sequence of planet building.

Because this feature causes motion both into and out of the NC and CC reservoirs, we argue that the feature was located between the reservoirs. Therefore, the feature that drove these isotopic evolutions and triggered planetesimal formation was likely the same feature that acted as the barrier during the first $\sim$0.95 Myr of the disk. The isotopic evolutions identified in this study provide multiple insights into the behaviour and evolution of this feature (Fig. \ref{schem}):
\begin{enumerate}
    \item It must prevent transport of material across it during the first $\sim$0.95 Myr, as evidenced by the pronounced difference in isotopic concentration of the NC and CC reservoirs at $\sim$0.95 Myr after CAI formation. 
    \item It must not promote planetesimal formation during this early time window. 
    \item It must concentrate and store CAIs during this period. 
    \end{enumerate}
The behaviour of this feature then changed at $\sim$0.95 Myr, after which:
\begin{enumerate}
    \item It started transporting NC and CC material past it in both directions, with a size dependence to inward transport.
    \item It created environments on both sides that were conducive to planetesimal formation.
    \item It caused CAIs to move from its radial distance, but overwhelmingly outwards.
\end{enumerate}
This feature, therefore, had several pronounced impacts on the behaviour of the disk and on planetesimal formation, and these effects changed drastically and abruptly.

By initiating planetesimal formation, this phenomenon played a key role the progressive series of events that formed the terrestrial planets. As such, identifying and understanding this feature is key to deciphering a holistic model of the transformation of dust into planets. We will explore this in future research by modelling different candidates (e.g., Jupiter, condensation lines, disk expansion) and assessing their ability to recreate the trends and behaviours we have presented.

\section{Conclusions}

An integral process during the terrestrial planet-building hierarchy is the formation of planetesimals. To better understand this process, we constructed a refined planetesimal formation age model. This model utilises recent advancements in nucleosynthetic isotopic compositions to produce the most complete accretion age records yet of these planetary building blocks. We find that NC and CC planetesimals started forming at very similar times, $\sim$0.95 Myr after CAI formation. The parent bodies of NC and CC iron meteorites continued forming over overlapping periods, while the parent bodies of NC and CC chondrites formed later during non-overlapping periods. By coupling these ages with the proportions of isotopic end-members we recovered from mixing models, we constructed records of the motion of material throughout the protoplanetary disk. These records indicate that both NC and CC material crossed the barrier in the disk. This process started at a very similar time as planetesimal formation, demonstrating that the feature that drove this motion also initiated planetesimal accretion. We argue this phenomenon was the semi-permeable barrier in the disk, and describe several effects it had on disk evolution and planetesimal formation. Despite several candidates having been proposed, the feature that served as this semi-permeable barrier is currently unknown. The behaviours and trends we present provide several novel constraints that can now be utilised in models to identify this feature. Because this feature caused planetesimal formation, it played a key role in terrestrial planet building. As such, identifying and understanding this feature is essential to uncovering a holistic model of the transformation of dust into planets.

\section{Acknowledgements}

JFJB and JPT acknowledge funding from the UKRI Research Frontier Guarantee program EP/Y014375/1. SS acknowledges funding through STFC. HRS acknowledges funding on a Natural Environment Research Council studentship NE/S007474/1, an Exonian Graduate Scholarship from Exeter College, University of Oxford and funding from the Research Council of Norway through the Centres of Excellence funding scheme, project number 332523 (PHAB). The authors thanks Tim Elliott for fruitful discussions. Prepared by LLNL under Contract DE-AC52-07NA27344.



\newpage

\begin{sidewaystable}
\begin{centering}
\begin{tabular}{|c|c|c|c|c|c|c|c|c|c|c|c|}
\hline 
 &  & Group & $T_{\mm{a}}$ (Myr) & +2\ensuremath{\sigma} (Myr) & -2\ensuremath{\sigma} (Myr) & $P_{\mm{CI}}$ & +2\ensuremath{\sigma} & -2\ensuremath{\sigma} & $P_{\mm{CAI}}$ & +2\ensuremath{\sigma} & -2\ensuremath{\sigma}\tabularnewline
\hline 
\hline 
\multirow{11}{*}{\begin{turn}{90}
NC
\end{turn}} & \multirow{5}{*}{\begin{turn}{90}
Irons
\end{turn}} & IC & 0.98 & 0.35 & 0.54 &  &  &  &  &  & \tabularnewline
\cline{3-12} \cline{4-12} \cline{5-12} \cline{6-12} \cline{7-12} \cline{8-12} \cline{9-12} \cline{10-12} \cline{11-12} \cline{12-12} 
 &  & IIAB & 1.07 & 0.29 & 0.44 & 0.04 & 0.05 & 0.05 &  &  & \tabularnewline
\cline{3-12} \cline{4-12} \cline{5-12} \cline{6-12} \cline{7-12} \cline{8-12} \cline{9-12} \cline{10-12} \cline{11-12} \cline{12-12} 
 &  & IIIAB & 1.30 & 0.21 & 0.37 & 0.04 & 0.03 & 0.03 &  &  & \tabularnewline
\cline{3-12} \cline{4-12} \cline{5-12} \cline{6-12} \cline{7-12} \cline{8-12} \cline{9-12} \cline{10-12} \cline{11-12} \cline{12-12} 
 &  & IVA & 1.39 & 0.18 & 0.36 & 0.23 & 0.03 & 0.03 &  &  & \tabularnewline
\cline{3-12} \cline{4-12} \cline{5-12} \cline{6-12} \cline{7-12} \cline{8-12} \cline{9-12} \cline{10-12} \cline{11-12} \cline{12-12} 
 &  & IIIE & 1.45 & 0.15 & 0.34 &  &  &  &  &  & \tabularnewline
\cline{2-12} \cline{3-12} \cline{4-12} \cline{5-12} \cline{6-12} \cline{7-12} \cline{8-12} \cline{9-12} \cline{10-12} \cline{11-12} \cline{12-12} 
 & \multirow{6}{*}{\begin{turn}{90}
Chondrites
\end{turn}} & H & 2.03 & 0.06 & 0.06 & 0.28 & 0.01 & 0.01 &  &  & \tabularnewline
\cline{3-12} \cline{4-12} \cline{5-12} \cline{6-12} \cline{7-12} \cline{8-12} \cline{9-12} \cline{10-12} \cline{11-12} \cline{12-12} 
 &  & L & 2.10 & 0.06 & 0.06 & 0.31 & 0.05 & 0.05 &  &  & \tabularnewline
\cline{3-12} \cline{4-12} \cline{5-12} \cline{6-12} \cline{7-12} \cline{8-12} \cline{9-12} \cline{10-12} \cline{11-12} \cline{12-12} 
 &  & LL & 2.08 & 0.06 & 0.06 & 0.26 & 0.03 & 0.03 &  &  & \tabularnewline
\cline{3-12} \cline{4-12} \cline{5-12} \cline{6-12} \cline{7-12} \cline{8-12} \cline{9-12} \cline{10-12} \cline{11-12} \cline{12-12} 
 &  & R & 1.86 & 0.05 & 0.05 & 0.45 & 0.03 & 0.03 &  &  & \tabularnewline
\cline{3-12} \cline{4-12} \cline{5-12} \cline{6-12} \cline{7-12} \cline{8-12} \cline{9-12} \cline{10-12} \cline{11-12} \cline{12-12} 
 &  & EH & 1.74 & 0.06 & 0.06 & 0.48 & 0.02 & 0.02 &  &  & \tabularnewline
\cline{3-12} \cline{4-12} \cline{5-12} \cline{6-12} \cline{7-12} \cline{8-12} \cline{9-12} \cline{10-12} \cline{11-12} \cline{12-12} 
 &  & EL & 1.98 & 0.06 & 0.06 & 0.48 & 0.02 & 0.02 &  &  & \tabularnewline
\hline 
\multirow{13}{*}{\begin{turn}{90}
CC
\end{turn}} & \multirow{5}{*}{\begin{turn}{90}
Irons
\end{turn}} & IID & 1.29 & 0.09 & 0.16 & 0.96 & 0.24 & 0.26 & 0.012 & 0.029 & 0.025\tabularnewline
\cline{3-12} \cline{4-12} \cline{5-12} \cline{6-12} \cline{7-12} \cline{8-12} \cline{9-12} \cline{10-12} \cline{11-12} \cline{12-12} 
 &  & IIC & 1.34 & 0.09 & 0.18 & 0.88 & 0.25 & 0.26 & 0.008 & 0.024 & 0.021\tabularnewline
\cline{3-12} \cline{4-12} \cline{5-12} \cline{6-12} \cline{7-12} \cline{8-12} \cline{9-12} \cline{10-12} \cline{11-12} \cline{12-12} 
 &  & IIIF & 1.44 & 0.17 & 0.41 & 0.75 & 0.25 & 0.27 & 0.016 & 0.027 & 0.023\tabularnewline
\cline{3-12} \cline{4-12} \cline{5-12} \cline{6-12} \cline{7-12} \cline{8-12} \cline{9-12} \cline{10-12} \cline{11-12} \cline{12-12} 
 &  & IIF & 1.77 & 0.14 & 0.24 & 0.64 & 0.14 & 0.14 & 0.040 & 0.023 & 0.021\tabularnewline
\cline{3-12} \cline{4-12} \cline{5-12} \cline{6-12} \cline{7-12} \cline{8-12} \cline{9-12} \cline{10-12} \cline{11-12} \cline{12-12} 
 &  & IVB & 1.61 & 0.08 & 0.16 & 0.63 & 0.13 & 0.14 & 0.013 & 0.013 & 0.012\tabularnewline
\cline{2-12} \cline{3-12} \cline{4-12} \cline{5-12} \cline{6-12} \cline{7-12} \cline{8-12} \cline{9-12} \cline{10-12} \cline{11-12} \cline{12-12} 
 & \multirow{8}{*}{\begin{turn}{90}
Chondrites
\end{turn}} & CK & 2.32 & 0.05 & 0.04 & 0.62 & 0.04 & 0.04 & 0.053 & 0.017 & 0.014\tabularnewline
\cline{3-12} \cline{4-12} \cline{5-12} \cline{6-12} \cline{7-12} \cline{8-12} \cline{9-12} \cline{10-12} \cline{11-12} \cline{12-12} 
 &  & CL & 2.55 & 0.07 & 0.07 & 0.73 & 0.03 & 0.03 & 0.025 & 0.004 & 0.004\tabularnewline
\cline{3-12} \cline{4-12} \cline{5-12} \cline{6-12} \cline{7-12} \cline{8-12} \cline{9-12} \cline{10-12} \cline{11-12} \cline{12-12} 
 &  & CV & 2.78 & 0.08 & 0.07 & 0.78 & 0.01 & 0.02 & 0.035 & 0.004 & 0.004\tabularnewline
\cline{3-12} \cline{4-12} \cline{5-12} \cline{6-12} \cline{7-12} \cline{8-12} \cline{9-12} \cline{10-12} \cline{11-12} \cline{12-12} 
 &  & CO & 2.79 & 0.08 & 0.07 & 0.75 & 0.06 & 0.06 & 0.038 & 0.012 & 0.011\tabularnewline
\cline{3-12} \cline{4-12} \cline{5-12} \cline{6-12} \cline{7-12} \cline{8-12} \cline{9-12} \cline{10-12} \cline{11-12} \cline{12-12} 
 &  & CM & 3.14 & 0.15 & 0.13 & 0.81 & 0.02 & 0.02 & 0.026 & 0.003 & 0.003\tabularnewline
\cline{3-12} \cline{4-12} \cline{5-12} \cline{6-12} \cline{7-12} \cline{8-12} \cline{9-12} \cline{10-12} \cline{11-12} \cline{12-12} 
 &  & TL & 3.55 & 0.28 & 0.20 & 0.89 & 0.04 & 0.04 & 0.017 & 0.007 & 0.006\tabularnewline
\cline{3-12} \cline{4-12} \cline{5-12} \cline{6-12} \cline{7-12} \cline{8-12} \cline{9-12} \cline{10-12} \cline{11-12} \cline{12-12} 
 &  & CR & 3.90 & 0.30 & 0.23 & 0.93 & 0.02 & 0.02 & 0.007 & 0.005 & 0.004\tabularnewline
\cline{3-12} \cline{4-12} \cline{5-12} \cline{6-12} \cline{7-12} \cline{8-12} \cline{9-12} \cline{10-12} \cline{11-12} \cline{12-12} 
 &  & CI & 2.70 & 0.13 & 0.13 & 1.00 & 0.03 & 0.03 & 0.00 & 0.002 & 0.002\tabularnewline
\hline 
\end{tabular}
\par\end{centering}
\caption{Calculated $T_{\mm{a}}$, $P_{\mm{CI}}$, and $P_{\mm{CAI}}$ for grouped
iron meteorites and chondrites from our model. Times are Myr after
CAI formation.\label{GroupedOutputs}}
\end{sidewaystable}

\begin{sidewaystable}
\begin{centering}
\begin{tabular}{|c|c|c|c|c|c|c|c|c|c|c|}
\hline 
 & Iron Meteorite & $T_{\mm{a}}$ (Myr) & +2\ensuremath{\sigma} (Myr) & -2\ensuremath{\sigma} (Myr) & $P_{\mm{CI}}$ & +2\ensuremath{\sigma} & -2\ensuremath{\sigma} & $P_{\mm{CAI}}$ & +2\ensuremath{\sigma} & -2\ensuremath{\sigma}\tabularnewline
\hline 
\hline 
\multirow{5}{*}{\begin{turn}{90}
NC ungrouped
\end{turn}} & EET 83230 & 0.14 & 0.69 & 0.97 &  &  &  &  &  & \tabularnewline
\cline{2-11} \cline{3-11} \cline{4-11} \cline{5-11} \cline{6-11} \cline{7-11} \cline{8-11} \cline{9-11} \cline{10-11} \cline{11-11} 
 & Reed City & 1.16 & 0.23 & 0.43 &  &  &  &  &  & \tabularnewline
\cline{2-11} \cline{3-11} \cline{4-11} \cline{5-11} \cline{6-11} \cline{7-11} \cline{8-11} \cline{9-11} \cline{10-11} \cline{11-11} 
 & Santiago Papasquiero & 1.33 & 0.21 & 0.74 &  &  &  &  &  & \tabularnewline
\cline{2-11} \cline{3-11} \cline{4-11} \cline{5-11} \cline{6-11} \cline{7-11} \cline{8-11} \cline{9-11} \cline{10-11} \cline{11-11} 
 & Washington County & 1.27 & 0.24 & 0.69 &  &  &  &  &  & \tabularnewline
\cline{2-11} \cline{3-11} \cline{4-11} \cline{5-11} \cline{6-11} \cline{7-11} \cline{8-11} \cline{9-11} \cline{10-11} \cline{11-11} 
 & Zacatecas (1792) & 0.76 & 0.41 & 0.69 &  &  &  &  &  & \tabularnewline
\hline 
\multirow{16}{*}{\begin{turn}{90}
CC ungrouped
\end{turn}} & Babb's Mill (Troost's Iron){*} & 1.18 & 0.55 & 0.87 & 0.55 & 0.24 & 0.25 & 0.042 & 0.041 & 0.034\tabularnewline
\cline{2-11} \cline{3-11} \cline{4-11} \cline{5-11} \cline{6-11} \cline{7-11} \cline{8-11} \cline{9-11} \cline{10-11} \cline{11-11} 
 & Grand Rapids & 1.14 & 0.53 & 0.85 & 0.53 & 0.21 & 0.21 & 0.030 & 0.031 & 0.026\tabularnewline
\cline{2-11} \cline{3-11} \cline{4-11} \cline{5-11} \cline{6-11} \cline{7-11} \cline{8-11} \cline{9-11} \cline{10-11} \cline{11-11} 
 & Piñon & 1.18 & 0.56 & 0.87 & 0.49 & 0.18 & 0.18 & 0.038 & 0.030 & 0.025\tabularnewline
\cline{2-11} \cline{3-11} \cline{4-11} \cline{5-11} \cline{6-11} \cline{7-11} \cline{8-11} \cline{9-11} \cline{10-11} \cline{11-11} 
 & Tucson & 1.89 & 0.10 & 0.46 & 0.59 & 0.33 & 0.35 & 0.032 & 0.041 & 0.034\tabularnewline
\cline{2-11} \cline{3-11} \cline{4-11} \cline{5-11} \cline{6-11} \cline{7-11} \cline{8-11} \cline{9-11} \cline{10-11} \cline{11-11} 
 & Mbosi & 1.50 & 0.37 & 0.74 & 0.56 & 0.33 & 0.35 & 0.034 & 0.044 & 0.036\tabularnewline
\cline{2-11} \cline{3-11} \cline{4-11} \cline{5-11} \cline{6-11} \cline{7-11} \cline{8-11} \cline{9-11} \cline{10-11} \cline{11-11} 
 & New Baltimore & 1.14 & 0.10 & 0.28 & 0.92 & 0.22 & 0.23 & -0.004 & 0.031 & 0.025\tabularnewline
\cline{2-11} \cline{3-11} \cline{4-11} \cline{5-11} \cline{6-11} \cline{7-11} \cline{8-11} \cline{9-11} \cline{10-11} \cline{11-11} 
 & Nordheim & 1.58 & 0.13 & 0.38 & 0.69 & 0.20 & 0.21 & 0.016 & 0.026 & 0.023\tabularnewline
\cline{2-11} \cline{3-11} \cline{4-11} \cline{5-11} \cline{6-11} \cline{7-11} \cline{8-11} \cline{9-11} \cline{10-11} \cline{11-11} 
 & NWA 6932 & 1.62 & 0.27 & 0.61 & 0.62 & 0.14 & 0.15 & 0.036 & 0.035 & 0.028\tabularnewline
\cline{2-11} \cline{3-11} \cline{4-11} \cline{5-11} \cline{6-11} \cline{7-11} \cline{8-11} \cline{9-11} \cline{10-11} \cline{11-11} 
 & Tishomingo & 1.58 & 0.12 & 0.44 & 0.67 & 0.20 & 0.21 & 0.013 & 0.026 & 0.022\tabularnewline
\cline{2-11} \cline{3-11} \cline{4-11} \cline{5-11} \cline{6-11} \cline{7-11} \cline{8-11} \cline{9-11} \cline{10-11} \cline{11-11} 
 & ALHA 77255 & 1.53 & 0.17 & 0.41 & 0.66 & 0.20 & 0.21 & 0.016 & 0.028 & 0.024\tabularnewline
\cline{2-11} \cline{3-11} \cline{4-11} \cline{5-11} \cline{6-11} \cline{7-11} \cline{8-11} \cline{9-11} \cline{10-11} \cline{11-11} 
 & Guffey & 1.45 & 0.14 & 0.35 & 0.73 & 0.26 & 0.27 & 0.011 & 0.032 & 0.027\tabularnewline
\cline{2-11} \cline{3-11} \cline{4-11} \cline{5-11} \cline{6-11} \cline{7-11} \cline{8-11} \cline{9-11} \cline{10-11} \cline{11-11} 
 & Hammond & 1.27 & 0.15 & 0.40 & 0.91 & 0.16 & 0.17 & 0.009 & 0.024 & 0.020\tabularnewline
\cline{2-11} \cline{3-11} \cline{4-11} \cline{5-11} \cline{6-11} \cline{7-11} \cline{8-11} \cline{9-11} \cline{10-11} \cline{11-11} 
 & ILD 83500{*} & 0.92 & 0.59 & 0.90 & 0.68 & 0.18 & 0.18 & 0.033 & 0.040 & 0.032\tabularnewline
\cline{2-11} \cline{3-11} \cline{4-11} \cline{5-11} \cline{6-11} \cline{7-11} \cline{8-11} \cline{9-11} \cline{10-11} \cline{11-11} 
 & Illinois Gulch & 1.11 & 0.18 & 0.55 & 0.89 & 0.19 & 0.20 & -0.0003 & 0.026 & 0.022\tabularnewline
\cline{2-11} \cline{3-11} \cline{4-11} \cline{5-11} \cline{6-11} \cline{7-11} \cline{8-11} \cline{9-11} \cline{10-11} \cline{11-11} 
 & La Caille & 1.40 & 0.15 & 0.42 & 0.76 & 0.20 & 0.20 & 0.009 & 0.023 & 0.020\tabularnewline
\cline{2-11} \cline{3-11} \cline{4-11} \cline{5-11} \cline{6-11} \cline{7-11} \cline{8-11} \cline{9-11} \cline{10-11} \cline{11-11} 
 & N'Goureyma & 1.35 & 0.10 & 0.48 & 0.85 & 0.11 & 0.11 & 0.005 & 0.024 & 0.020\tabularnewline
\hline 
\end{tabular}
\par\end{centering}
\caption{Calculated $T_{\mm{a}}$, $P_{\mm{CI}}$, and $P_{\mm{CAI}}$ for ungrouped
iron meteorites. Times are Myr after CAI formation. {*} denotes part
of the South Byron Trio.\label{UngroupedOutputs}}
\end{sidewaystable}

\newpage
\clearpage


  \bibliographystyle{elsarticle-harv} 
  \bibliography{lib.bib}







\end{document}